\documentstyle[12pt]{article}
\begin{document}

\begin{flushright}
EDO-EP-49\\
DFPD 05/TH/16\\
June, 2005\\
\end{flushright}
\vspace{20pt}

\pagestyle{empty}
\baselineskip15pt

\begin{center}
{\large\bf Y-formalism in Pure Spinor Quantization of
Superstrings
\vskip 1mm
}

\vspace{20mm}

Ichiro Oda
          \footnote{
          E-mail address:\ ioda@edogawa-u.ac.jp
                  }
\\

\vspace{5mm}
          Edogawa University,
          474 Komaki, Nagareyama City, Chiba 270-0198, Japan\\

\vspace{5mm}

and

\vspace{5mm}

Mario Tonin
          \footnote{
          E-mail address:\ mario.tonin@pd.infn.it
                  }
\\
\vspace{5mm}
          Dipartimento di Fisica, Universita degli Studi di Padova,\\
          Instituto Nazionale di Fisica Nucleare, Sezione di Padova,\\
          Via F. Marzolo 8, 35131 Padova, Italy\\

\end{center}


\vspace{5mm}
\begin{abstract}
We present the Y-formalism in the pure spinor quantization of 
superstring theory in detail.
Even if the $\omega-\lambda$ OPE is not completely free owing to the presence
of the projector reflecting the pure spinor constraint, it is shown that one 
can construct at the quantum level a conformal field theory in full
agreement with the Berkovits' pure spinor formalism. The OPE's of the two 
formalisms are the same if we define the relevant operators involving $\omega$
in a suitable manner. Moreover the 
Y-formalism at hand is utilized to find the full expression of the covariant, 
picture raised $b$ antighost $b_B$,  with a reasonable amount of effort.
\end{abstract}

\newpage
\pagestyle{plain}
\pagenumbering{arabic}

\section{Introduction}

The new approach of Berkovits \cite{Ber1}-\cite{Ber9} based on pure spinors 
has solved a well-known old problem of super Poincar\'e covariant quantization 
of superstrings. This approach is based on a critical action involving the 
Green-Schwarz (GS) matter fields and a couple (or two couples for type II 
superstrings) of conjugate, twistor-like ghost fields 
$\omega_\alpha$ and $\lambda^\alpha$ where $\lambda^\alpha $ satisfies the 
pure spinor constraint  $(\lambda \Gamma^a \lambda) = 0$. The BRST operator
 $Q$ is 
constructed out of the pure spinor $\lambda$ and the GS fermionic constraint 
and is nilpotent due to the pure spinor constraint. Then it turns out that the
cohomology of $Q$ reproduces the correct superstring spectrum \cite{Ber3}. 
Moreover, it has been shown that the pure spinor BRST operator $Q$ is related by 
a similarity transformation to the BRST operator of the Neveu-Schwarz-Ramond
(NSR) \cite{Ber4} and that $Q$ is also related by a similarity transformation to the
BRST operator of the GS superstring in light-cone gauge \cite{Ber7, Kazama4}.

The method to compute tree amplitudes in the pure spinor approach has been 
proposed and used to compute tree amplitudes in complete agreement with
that of the NRS formulation \cite{Ber2, Ber4}. However, to extend this method 
to loop amplitudes is not straightforward at all. The reason is that the $b$ 
antighost field, which is needed to compute higher loop amplitudes, does not appear 
naturally in this formalism due to the absence of diffeomorphism ghosts and, 
even as a compound field, it cannot be written 
without breaking super-Poincar\'e invariance.  Nevertheless, in a recent 
important paper \cite{Ber6}, Berkovits has presented a consistent recipe for 
computing loop amplitudes so that we can now evaluate  loop amplitudes at least in 
principle. This recipe is based  on  three ingredients: a Lorentz-invariant measure 
factor for the ghosts, suitable picture changing operators (PCO's) to absorb the zero 
modes of the ghosts and $(3g-3)$ (1 at genus 1) insertions of a picture raised 
$b_B$ field whose anticommutator with $Q$ gives the stress-energy tensor $T$ 
times a picture changing operator $Z_B$.
A clear discussion of this recipe is given in \cite{Sch} where the pure spinor 
approach is applied to the construction of D-brane boundary states. (For related 
constructions, see also \cite{Ang1, Muk}.)
         
The pure spinor approach has also been worked for superparticles in 10 
dimensions \cite{Ber9} and in 11 dimensions  \cite{Ber10}  
as well as for the supermembrane \cite{Ber10}, even if at present
in the last case only at the classical level.
A BRST approach for superparticles has been proposed in \cite{Chest}.
A recipe to compute one loop and multiloop amplitudes  for superparticles
in  d=10 and, interestingly, in d=11 has been worked out in 
\cite{Anguelova, Grassi3}. Moreover
$\sigma$-models \cite{Ber11, Oda1}  and superstrings in $AdS_5 \times S^5$
\cite{Ber12} has been formulated in this approach as well. In particular it 
has been shown \cite{Ber11} that BRST invariance  reproduces 
the correct field equations of the background fields.   

The pure spinor formulation and the GS one share with a manifest 
super-Poincar\'e invariance at the classical level, but the $\kappa$ symmetry 
in the GS superstring prevents a covariant quantization of it, so that only 
light-cone quantization is available for the GS superstring. This leads to a
complication due to the insertion of non-covariant contact vertex operators 
and also to an impossibility to work in backgrounds that do not admit 
a light-like Killing vector. In the NSR case, super-Poincar\'e invariance is 
reached only after summing over spin structures and the price to pay is surface 
terms at the boundary of the moduli space, which have to be handled carefully. 
Moreover the spin fields make it difficult to compute amplitudes with external fermions 
and on backgrounds with R-R fields.

It is worthwhile to stress that the pure spinor formulation is free from all 
these problems. Accordingly, the pure spinor formalism appears today as a new,
very promising, consistent formulation of superstring theories, an alternative 
to the NSR and GS ones, that shares the advantages of these two
formulations without suffering from their disadvantages. 

The statement that the pure spinor approach provides a super-Poincar\'e
covariant quantization of superstring theories is correct but deserves a 
warning. Besides an intrinsic non-covariance present in PCO's which however
is common to all systems with non-Lorentz scalar zero modes and is harmless, 
there is a hidden source of non-covariance and/or non-free fields at
intermediate steps, which is related to the pure spinor constraint.
What happens here is that the pure spinor constraint can be considered as the 
generator of a local gauge symmetry involving (only) $\omega$ (that we will refer to
$\omega$-symmetry) and this symmetry has to be gauge fixed in some way.

One strategy is to remove the pure spinor constraint. In \cite{Grassi1, 
Grassi2} new ghosts and ghosts of ghost are added in order to still have a 
nilpotent BRST operator, but then one is forced to have in the action terms which are
not quadratic in the fields. In \cite{Kazama1}-\cite{Kazama3} a more standard 
BRST-BV gauge fixing procedure is performed by adding ghosts and ghosts of ghost that 
break the Wick rotated, Lorentz group $SO(10)$ to $U(5)$.

The other strategy is to insist on preserving the pure spinor constraint.
Then problems with super-Poincar\'e covariance do not arise if only gauge 
invariant quantities with respect to $\omega$-symmetry, are involved.
However, at intermediate stages, (for instance in order to compute OPE's
between gauge invariant quantities) one needs to work with non-gauge invariant 
quantities and as a result the violation of super-Poincar\'e covariance is 
unavoidable. 
Also, as already noted, the $b$ field, which has ghost number $-1$, cannot be 
written in a covariant way. 

In \cite{Ber1, Ber3} the pure spinor constraint is
 resolved, thereby breaking the Wick rotated, Lorentz group $SO(10)$
to $U(5)$ and a $U(5)$-formalism is used not only to compute OPE's between 
gauge invariant quantites but also to write an expression for the 
(non-covariant) $b$ field.  
A natural way to deal with this unavoidable non-covariance arises in the 
approach of \cite{Matone} where a geometrical interpretation of the pure spinor
formulation, based on a twisted and gauge fixed superembedding approach, 
has been proposed. Here a non-covariant, spinor-like field $Y_\alpha$
possesses the intrinsic non-covariance of non-gauge invariant field 
with respect to $\omega$-symmetry and of the $b$ field. This Y-formalism, 
closely relating to the $U(5)$-formalism, turns out to be useful to compute 
OPE's and to deal with the $b$ field.

The aim of this paper is to develop the Y-formalism in a more systematic
way than in \cite{Matone}. In particular, in section 2, we will review the 
pure spinor approach from the point of view of \cite{Matone} in order to 
introduce this 
formalism and to give the expression of the $b$ field that arises
naturally in this framework. In section 3 we use the formalism to 
compute the OPE's involving the $\omega$ field. One surprise here is that
even if the $\omega-\lambda$ OPE is not free because of the projection
operator relevant to the pure spinor constraint, we can obtain a 
self-consistent conformal field theory which at the quantum level reproduces 
the same OPE's of the Berkovits' pure spinor formalism.
Section 4 is devoted to discuss the $b$ field and to study its properties. As 
mentioned above, an important ingredient in the Berkovits' method \cite{Ber6} 
for computing loop amplitudes is a picture raised $b$ field, 
denoted as $b_B$, of ghost number $0$, whose 
anticommutator with $Q$ is $TZ_B$. In \cite{Oda2} it has been shown that the 
$b$ field obtained in \cite{Ber3, Matone} if 
associated with $Z_B$, is equivalent to $b_B$ up to a BRST-exact term.
However the full expression of $b_B$ has not been given explicitly in \cite
{Ber6, Oda2}. Here, after reviewing the relation between $b$ and $b_B$, we use
the Y-formalism to recover the full expression of $b_B$.

\section{Pure spinor formulation and Y-formalism}

In this section, we start with a brief review of the pure spinor formalism of  
superstrings and then explain the Y-formalism.
For simplicity, let us consider only the left-moving (holomorphic) 
sector of a closed superstring theory. The pure spinor approach is based on 
the BRST charge
\begin{eqnarray}
Q = \oint (\lambda^\alpha d_\alpha),
\label{2.1}
\end{eqnarray}
and the action 
\begin{eqnarray}
I = \int (\frac{1}{2} \partial X^a \bar \partial X_a + p_\alpha
\bar \partial \theta^\alpha + \omega_\alpha \bar \partial \lambda^\alpha),
\label{2.2}
\end{eqnarray}
or equivalently \'a la Siegel \cite{Siegel},
\begin{eqnarray}
I= \int \Big[\frac{1}{2} \Pi^a \bar \Pi_a + \frac{1}{4} \Big( \Pi^a 
(\theta \Gamma_a \bar \partial \theta) - \bar \Pi^a (\theta \Gamma_a 
\partial \theta) \Big)+ d_\alpha \bar \partial \theta^\alpha + \omega_\alpha
\bar \partial \lambda^\alpha \Big],
\label{2.3}
\end{eqnarray}
where $\lambda$ is a pure spinor satisfying the equation
\begin{eqnarray}
(\lambda \Gamma^a \lambda) = 0,
\label{2.4}
\end{eqnarray}
and $d_\alpha$ and $\Pi^a$ are defined as
\begin{eqnarray}
d_\alpha &=& p_\alpha - \frac{1}{2}(\partial X^a + \frac{1}{4} \theta
\Gamma^a \partial \theta) (\Gamma_a \theta)_\alpha, \nonumber\\
\Pi^a &=& \partial X^a + \frac{1}{2} (\theta \Gamma^a \partial \theta), \nonumber\\
\bar \Pi^a &=& \bar \partial X^a + \frac{1}{2} (\theta \Gamma^a \bar \partial \theta).
\label{2.5}
\end{eqnarray}
This action is manifestly invariant under (global) super-Poincar\'e 
transformations; $\omega_{\alpha}$ is the conjugate momentum of the pure 
spinor ghost
$\lambda^\alpha$  and $\Pi^a$ and $d_\alpha$ are the super-Poincar\'e 
covariant momenta of the superspace coordinates  
$Z^{M} =( X^a, \theta^\alpha)$. In particular, note that $\Pi^a$ and 
$d_\alpha$ (anti)commute with the charges of target supersymmetry with 
currents given by
\begin{eqnarray}
q_\alpha = p_\alpha + \frac{1}{2}(\partial X^a 
+ \frac{1}{12} \theta \Gamma^a \partial \theta) (\Gamma_a \theta)_\alpha,
\label{2.6}
\end{eqnarray}
  
It is easily shown that the action $I$ is invariant under 
the BRST transformation generated by the BRST charge $Q$
which is nilpotent owing to the pure spinor constraint (\ref{2.4}).  
The cohomology of $Q$ defines the physical spectrum which is shown to be
equivalent to that of the RNS formalism and of the Green-Schwarz formalism
 in the light-cone gauge. Notice that in order to use $Q$ 
as BRST charge it is implicit that the pure spinor constraint is required 
to vanish in a strong sense.

Moreover, the action $I$ is invariant under the $\omega$-symmetry
\begin{eqnarray}
\delta \omega_\alpha = \Lambda_a (\Gamma^a \lambda)_\alpha.
\label{2.7}
\end{eqnarray}
where $\Lambda^a$ are local gauge parameters.
At the classical level the ghost current is of form
\begin{eqnarray}
J= (\omega \lambda),
\label{2.8}
\end{eqnarray}
and the Lorentz current for the ghost sector is given by
\begin{eqnarray}
N^{ab} = {1 \over 2} (\omega \Gamma^{ab} \lambda),
\label{2.9}
\end{eqnarray}
which together with $(\omega \partial \lambda)$ are not only the
 super-Poincar\'e 
covariant fields involving $\omega$ but also gauge invariant under 
$\omega$-symmetry.

{}From the field equations, we find that $p$, $\theta$, $\omega$ and $\lambda$ 
are holomorphic fields and at the quantum level, starting from the free field 
equal-time commutators for $X$ and the system $p-\theta$ one obtains 
the following OPE's involving $\Pi^a$ , $d_\alpha$ and the supercoordinates
\begin{eqnarray}
X^a(y) X^b(z) &\rightarrow& - \eta^{ab} \log(y - z), \nonumber\\
p_\alpha(y) \theta^\beta(z) &\rightarrow& \frac{1}{y - z} \delta_\alpha^\beta,
 \nonumber\\
d_\alpha(y) d_\beta(z) &\rightarrow& - \frac{1}{y - z} \Gamma^a_{\alpha\beta} 
\Pi_a(z), \nonumber\\
d_\alpha(y) \Pi^a(z) &\rightarrow& \frac{1}{y - z} (\Gamma^a 
\partial \theta)_\alpha(z).
\label{2.10}
\end{eqnarray}

As for the ghost sector, it would be inconsistent to assume a free field OPE
between $\omega$ and $\lambda$. Indeed in the ghost sector the action $I$ 
is not truly free.  The reason is that since the pure spinor constraint 
must vanish identically, not all the components of $\lambda$ are independent:
solving the constraint, five of them are expressed nonlinearly in terms of 
the others. Accordingly, five components of $\omega$ are pure gauge (this is 
the $\it{ 'reason \quad  d'etre'}$ of the $\omega$-symmetry). Notice that, 
with the free field OPE for $\omega$ and $\lambda$, $Q$ is not truly nilpotent
since when we operate $Q^2$ on $\omega$, $Q^2 \omega$ reproduces the gauge 
transformation 
(\ref{2.7}) with the gauge parameter $\Lambda^a = \Pi^a$. 
This inconsistency can be also seen from the fact that since the pure spinor 
constraint vanishes strongly, its OPE with any field and in particular with 
$\omega$ 
should vanish, which is not the case with the free field $\omega-\lambda$ OPE.

This difficulty can be overcome as follows.
Let us define the non-covariant object 
\begin{eqnarray}
Y_\alpha = \frac{v_\alpha}{(v \lambda)},
\label{2.11}
\end{eqnarray}
such that
\begin{eqnarray}
(Y \lambda) = 1,
\label{2.12}
\end{eqnarray}
where $v_\alpha$ is a constant pure spinor satisfying 
\begin{eqnarray}
(Y \Gamma^a Y) = 0.
\label{2.13}
\end{eqnarray}
Then it is useful to define the projector
\begin{eqnarray}
K_\alpha \ ^\beta= \frac{1}{2}(\Gamma^a \lambda)_\alpha
(Y \Gamma_a)^\beta,
\label{2.14}
\end{eqnarray}
which projects on a 5 dimensional subspace of the 16 dimensional spinor 
space in ten dimensions since we have $Tr K = 5$. Note that the orthogonal 
projector is given by $(1 - K)_\alpha \ ^\beta$. 
Now the pure spinor constraint implies
\begin{eqnarray}
\lambda^\alpha K_\alpha \ ^\beta = 0.
\label{2.15}
\end{eqnarray}
Since $K$ projects on a 5 dimensional subspace, Eq. (\ref{2.15}) is a simple way 
to understand why a pure spinor has eleven independent components.
Another interesting identity is
\begin{eqnarray}
( ( 1-K ) \Gamma^a \lambda)= 0.
\label{2.16}
\end{eqnarray}

Now we postulate the following OPE between $\omega$ and $\lambda$:
\begin{eqnarray}
\omega_\alpha (y) \lambda^\beta (z) \rightarrow \frac{1}{y-z} 
(\delta_\alpha^\beta - K_\alpha \ ^\beta (z)).
\label{2.17}
\end{eqnarray}
It follows from Eq. (\ref{2.17}) that the OPE between $\omega$ and
the pure spinor constraint vanishes identically. Moreover, the BRST charge
$Q$  is then strictly nilpotent even acting on $\omega$.
It is useful to notice that, with the help of the projector $K$, one can 
obtain a non-covariant but gauge-invariant antighost $\tilde\omega$ defined as
\begin{eqnarray}
\tilde \omega_\alpha = ( 1 - K )_\alpha \ ^\beta \omega_\beta, 
\label{2.18}
\end{eqnarray}
with the OPE
\begin{eqnarray}
\tilde \omega_\alpha (y) \lambda^\beta (z) \rightarrow \frac{1}{y-z}
( 1 - K(y) )_\alpha \ ^\gamma ( 1 - K(z) )_\gamma \ ^\beta.
\label{2.19}
\end{eqnarray}
As we shall show in next section, Eq. (\ref{2.17}) allows us to compute correctly
all the relevant OPE's.

The Y-formalism is also useful to deal with the $b$ field. Recall that 
the $b$ field plays an important role in computing higher loop amplitudes.  Its main property is
\begin{eqnarray}
\{ Q, b(z) \} = T(z),
\label{2.20}
\end{eqnarray}
where $T$ is the stress energy tensor which at the classical level is given by
\begin{eqnarray}
T = - \frac{1}{2} \Pi^a \Pi_a - d_\alpha \partial \theta^\alpha
+ \omega_\alpha \partial \lambda^\alpha.
\label{2.21}
\end{eqnarray}
Since in the pure spinor formulation the reparametrization ghosts do not
exist, $b$ must be a compound field. Moreover since the $b$ ghost has ghost 
number $-1$ and the fields which include $\omega$ and are gauge-invariant 
under the $\omega$ symmetry, always have ghost number zero or positive, one 
must use $Y_\alpha$ (which also has ghost number $-1$) to construct it. 
Therefore $b$ is not super-Poincar\'e invariant.  The $b$ ghost has been 
constructed for the first time in \cite{Ber3} in the U(5) formalism in such a way 
that it satisfies Eq. (\ref{2.20}). In the Y-formalism at hand
the $b$ ghost is
\begin{eqnarray}
b = \frac{1}{2} (Y \Pi^a \Gamma_a d) + (\omega ( 1 - K ) \partial \theta).
\label{2.22}
\end{eqnarray}

This expression of the $b$ ghost can be understood quite naturally in the 
superembedding approach to the pure spinor formulation in \cite{Matone} 
even if this approach was formulated only for the case of the heterotic string.
In the superembedding approach \cite{Dima}, strings are described by the embedding of 
an $n$ extended world-sheet supersymmetry in the target superspace ($n\le 8$), 
which is restricted by the condition that the pull-back  of the vector
 supervielbeins along the odd directions of the cotangent
 world-sheet superspace  vanishes. 
At the classical level, the superembedding approach is equivalent to the 
Green-Schwarz one and the $n$ world-sheet supersymmetries replace $n$ components 
of the $\kappa$ symmetry.
The connection between the superembedding approach and the pure spinor formalism
was obtained in \cite{Matone}, by starting with a complexified version of the $n=2$ 
superembedding model \cite{Tonin, Berk} and then performing twisting and suitable gauge fixing.
A peculiar feature in this model is that $X^a$ and $\theta^\alpha$ are complex 
$n=2$ world-sheet superfields.$\footnote{The fact that they are complex does not
double the dynamical degrees of freedom since their complex conjugates are not 
present in the action.}$
Then, the superembedding constraint fixes all the higher components of the
superfield $X^a$ and gives rise to suitable constraints to the higher 
components of $\theta^\alpha$ which are a couple of twistor-like spinors,
called $\lambda$ and $\bar \lambda$ with ghost number 1 and $-1$, respectively 
and an anti-commuting spinor called $\sigma^\alpha$.
In particular, it has been shown that the constraints for
$\lambda$ and  $\bar \lambda$ are nothing but the pure spinor constraint
\begin{eqnarray}
(\lambda \Gamma^a \lambda) = 0 = (\bar \lambda \Gamma^a \bar \lambda),
\nonumber
\end{eqnarray}
in addition to
\begin{eqnarray}
(\lambda \Gamma^a \bar \lambda) = \Pi^a.
\label{2.23}
\end{eqnarray}

Before twisting, the model has an $n=2$ superfield of currents whose components 
are the U(1) current of ghost number, the two currents of the supersymmetry 
charges $Q$ and $\bar Q$, and the stress energy tensor.
The action of $Q$ and $\bar Q$ on $\theta$, $\lambda$ and $\bar \lambda$ is
given by
\begin{eqnarray}
Q \theta &=& \lambda, \nonumber\\
\bar Q \theta &=& \bar \lambda, \nonumber\\
\bar Q \lambda &=& \partial \theta + \sigma, \nonumber\\
Q \bar \lambda &=& \partial \theta - \sigma. 
\label{2.24}
\end{eqnarray}
The procedure of twisting shifts the weight of $\lambda$ and $\bar \lambda$ 
(both have the weight $1/2$ before twisting) to 
$0$ and $1$, respectively and at the same time the two currents of $n=2$ 
supersymmetry $Q$ and $\bar Q$ become to have the weight $1$ and $2$, respectively.
Thus, $Q$ can be identified as a BRST charge while the current of the second 
supersymmetry (with weight $2$) is (two times) the $b$ field.
It is interesting to note that twisting also shifts the central charge of 
the system of superdiff. ghosts from $6$ to zero, which explains why the 
superdiff. ghosts do not arise in the pure spinor approach.

The model is also invariant under $\kappa$-symmetry with $6$ superfield 
components. By gauge fixing the $\kappa$-symmetry relevant to 
$\bar \lambda^\alpha$ and $\sigma^\alpha$ and taking account of the constraint of
$\bar \lambda$ (i.e., the pure spinor constraint and Eq. (\ref{2.23})) and of $\sigma$,
one can specify completely these fields in terms of the other fields and $Y_\alpha$.  
Actually we have 
\begin{eqnarray}
\bar \lambda^\alpha = (\Gamma^a \Pi_a Y)^\alpha,
\label{2.25}
\end{eqnarray}
and
\begin{eqnarray}
\sigma^\alpha = (( 1 -2 \bar K) \partial \theta)^\alpha.
\label{2.26}
\end{eqnarray}
As mentioned above, after this partial gauge fixing procedure, the currents of $Q$
and $\bar Q$ are of form
\begin{eqnarray}
\hat j = \lambda^\alpha d_\alpha,
\label{2.27}
\end{eqnarray}
and
\begin{eqnarray}
2b = \bar \lambda^\alpha d_\alpha + \tilde \omega_\alpha
(\partial \theta^\alpha + \sigma^\alpha).
\label{2.28}
\end{eqnarray}
(A term similar to the second one in (\ref{2.28}), involving the conjugate 
momentum of $\bar \lambda$, does not arise in (\ref{2.27}) and terms involving
 the conjugate 
momentum of $\sigma$  do no appear in (\ref{2.27}) and (\ref{2.28}) 
since $\bar \lambda$ and $\sigma$ are not independent fields but are given in 
(\ref{2.25}) and (\ref{2.26}) ).

$\hat j$ means the current of the BRST charge $Q$ and from Eqs. (\ref{2.25}),
(\ref{2.26}) and (\ref{2.28}) one reproduces the $b$ ghost in Eq.
(\ref{2.22}).
To conclude this short review of \cite{Matone}, let us mention that the 
remaining $\kappa$-symmetry that acts on $\theta$ and $\lambda $ can be fixed
 \'a la BRST 
and as a result one can reproduce the pure spinor action from the action of 
the superembedding model via twisting and gauge fixing.
       
There seems to be a possible difficulty when we use the Y-formalism without care. 
Indeed, since one has a relation
\begin{eqnarray}
\lbrace Q,(Y \theta) \rbrace = (Y \lambda) = 1,
\label{2.29}
\end{eqnarray}
one might be afraid that the $Q$-cohomology at hand is trivial in that
any BRST-closed operator A can be written as
\begin{eqnarray}
A = \lbrace Q, (Y\theta) A \rbrace. 
\label{2.30}
\end{eqnarray}
A possible resolution of this problem can be found along the lines of Ref.
\cite{Ber3} where "the Almost
Super Poincar\'e Covariant operators" (ASPC operators) have been defined.
They are the operators which are covariant only under the subgroup of the 
super-Poincar\'e group that breaks $SO(10)$ but preserves its subgroup $U(5)$. 
The problem could be solved if one restricts the space of operators that 
define
the cohomology of $Q$, to ASPC operators. Indeed $(Y \theta)$ is not ASPC 
since it is not invariant under the action of the supersymmetry charges.   
However this problem requires further examination since there are interesting 
operators like picture changing operators for $\lambda$ zero modes or Lorentz
currents for matter fields which are not ASPC.

\section{Various OPE's including $\omega_\alpha$}

In order to show that our formalism using the non covariant $Y_\alpha$ field
is equivalent to that of Berkovits, it is necessary to prove that
several important OPE's such as that of the Lorentz current, the ghost current 
and the stress energy tensor are the same in the both formalisms. 
Thus, in this section, we shall calculate those OPE's explicitly.\footnote{ 
Some of the results of this section were first obtained in \cite{Luca}.}
It is worthwhile to notice that the nontrivial
difference between our formalism and the Berkovits one appears in the
OPE's including $\omega_\alpha$ (as well as $\lambda^\alpha$), so we
will focus on the following OPE's which have been already given by Berkovits 
\cite {Ber1, Ber3} (see also \cite {Ber13} where the central 
charges are also obtained in a manifestly covariant way)\footnote{
We use the convention  $ X^{(A}Y^{B)} = X^A Y^B + X^B Y^A$ and 
$ X^{[A} Y^{B]} = X^A Y^B - X^B Y^A $. } :
\begin{eqnarray}
T(y) T(z) \rightarrow \frac{2}{(y-z)^2} T(z) + \frac{1}{y-z} \partial T(z),
\label{3.1}
\end{eqnarray}
\begin{eqnarray}
T(y) J(z) \rightarrow \frac{8}{(y-z)^3} + \frac{1}{(y-z)^2} J(z) 
+ \frac{1}{y-z} \partial J(z),
\label{3.2}
\end{eqnarray}
\begin{eqnarray}
T(y) N^{ab}(z) \rightarrow \frac{1}{(y-z)^2}N^{ab}(z) + \frac{1}{y-z}
\partial N^{ab}(z),
\label{3.3}
\end{eqnarray}
\begin{eqnarray}
J(y) J(z) \rightarrow - \frac{4}{(y-z)^2},
\label{3.4}
\end{eqnarray}
\begin{eqnarray}
J(y) N^{ab}(z) \rightarrow 0,
\label{3.5}
\end{eqnarray}
\begin{eqnarray}
N^{ab}(y) N^{cd}(z) \rightarrow - \frac{3}{(y-z)^2} \eta^{d[a}\eta^{b]c}
- \frac{1}{y-z} (\eta^{a[c}N^{d]b} - \eta^{b[c}N^{d]a}).
\label{3.6}
\end{eqnarray}
If, in our formalism, the stress tensor $T$, the Lorentz currents $N^{ab}$, 
the ghost current $J$ are defined as in Eqs. (\ref{2.8}), (\ref{2.9}) and
(\ref{2.21}), spurious terms involving $Y_{\alpha}$ arise in their OPE's,
so it turns out that the definitions of $J$, $N^{ab}$
and $T$ must be changed. Indeed, since $\partial Y_{\alpha}$ has the 
same (naive) conformal weight as $\tilde \omega$, terms involving $\partial Y$ 
(and $Y$) can arise in these definitions \cite{Luca}.   
We have in fact found that the OPE's (\ref{3.1})-(\ref{3.6}) are also 
satisfied in our formalism involving the non-covariant field $Y_\alpha$ 
if we define the corresponding quantities as follows:
\begin{eqnarray}
T &=& - \frac{1}{2} \partial X^a \partial X_a - p_\alpha \partial \theta^\alpha
+ \omega_\alpha \partial \lambda^\alpha + \frac{3}{2} \partial(Y \partial 
\lambda)
\nonumber\\ 
&=& - \frac{1}{2} \Pi^a \Pi_a - d_\alpha \partial \theta^\alpha
+ \omega_\alpha \partial \lambda^\alpha + \frac{3}{2} \partial(Y \partial 
\lambda),
\label{3.7}
\end{eqnarray}
\begin{eqnarray}
N^{ab} = \frac{1}{2} \Big[ (\omega \Gamma^{ab} \lambda) 
+ \frac{1}{2} (\partial Y \Gamma^{ab} \lambda) - 2 \partial (Y \Gamma^{ab} 
\lambda) \Big],
\label{3.8}
\end{eqnarray}
\begin{eqnarray}
J = (\omega \lambda) - \frac{7}{2} (\partial Y \lambda).
\label{3.9}
\end{eqnarray}

In this section it is convenient, in order to avoid misunderstandings, 
to rename the ghost number and Lorentz currents in (\ref{2.8}), (\ref{2.9}), 
 and the stress energy tensor of the ghosts in (\ref{2.21}) as :
\begin{eqnarray}
J_0 &=& (\omega \lambda), \nonumber\\
N^{ab}_0 &=& \frac{1}{2}(\omega \Gamma^{ab} \lambda), \nonumber\\
T^{(\omega\lambda)}_0 &=& (\omega \partial \lambda),
\label{3.10}
\end{eqnarray}
and write
\begin{eqnarray}
T^{(\omega\lambda)} &=& T^{(\omega\lambda)}_0 + \frac{3}{2} \partial
(Y \partial \lambda),
\label{3.11}
\end{eqnarray}
where $T^{(\omega\lambda)}$ denotes the stress energy tensor of the ghosts.
Here let us note that Y-dependent terms in Eqs. (\ref{3.8}), (\ref{3.9})
and (\ref{3.11}) are BRST-exact in that they can be rewritten  as
\begin{eqnarray}
N^{ab} &=& N^{ab}_{0} + \{ Q, X_N \}, \nonumber\\
J &=& J_0 + \{ Q, X_J \}, \nonumber\\
T^{(\omega\lambda)} &=& T^{(\omega\lambda)}_0 + \{ Q, X_T \}, 
\label{3.12}
\end{eqnarray}
where $X_{N} = - (Y \Gamma^{ab} \partial \theta) + \frac{3}{4}(Y\Gamma^{ab}
\lambda)(Y\partial \theta)$, $ X_{J} = \frac{7}{2}(Y\partial\theta)$ and
$ X_{T} = \frac{3}{2} \partial(Y\partial \theta) $ are ASPC operators.
In order to compute the OPE's (\ref{3.1})-(\ref{3.6}), we should notice 
that 
the OPE between $\omega$ and $\lambda$
is not free owing to the existence of the projection $K$, so it is useful to 
make use of the generalized normal ordering whose definition is given by the
contour integration \cite{Francesco} 
\begin{eqnarray}
(A B)(z) = \oint_z \frac{dw}{w-z} A(w) B(z).
\label{3.13}
\end{eqnarray}
Then the OPE of $A(z)$ and $B(w)$ is described by
\begin{eqnarray}
A(z) B(w) = <A(z) B(w)> + (A(z) B(w)),
\label{3.14}
\end{eqnarray}
where $<A(z) B(w)>$ denotes the contraction containing $\it{all}$ the singular
terms of the OPE and $(A(z) B(w))$ stands for the complete sequence of 
regular terms whose explicit forms can be extracted from the Taylor
expansion of $A(z)$ around $w$:
\begin{eqnarray}
(A(z) B(w)) = \sum_{k \geq 0} \frac{(z-w)^k}{k!} (\partial^k A \cdot B)(w).
\label{3.15}
\end{eqnarray}
Moreover, we have to use the generalized Wick theorem given by
\begin{eqnarray}
<A(z)(B C)(w)> = \oint_w \frac{dx}{x-w} \Big[ <A(z) B(x)> C(w)
+ B(x) <A(z) C(w)> \Big].
\label{3.16}
\end{eqnarray}
{}From this definition, it is important to notice that the first $\it{regular}$
term of the various OPE's always contributes.

Let us start by evaluating the Lorentz algebras (\ref{3.6}). We can easily 
calculate the singular terms of $N_0$-$N_0$ OPE
\begin{eqnarray}
&{}& <N_0^{ab}(y) N_0^{cd}(z)> \nonumber\\
&=& \frac{1}{2} (\Gamma^{cd})^\alpha \ _\beta 
\oint_z \frac{dx}{x-z} \Big[ <N_0^{ab}(y) \omega_\alpha(x)> \lambda^\beta(z)
+ \omega_\alpha(x) <N_0^{ab}(y) \lambda^\beta(z)> \Big] \nonumber\\
&\equiv& I_1 + I_2.
\label{3.17}
\end{eqnarray}
We shall evaluate $I_1$ and $I_2$ in order.
\begin{eqnarray}
&{}& I_1 \equiv \frac{1}{2} (\Gamma^{cd})^\alpha \ _\beta 
\oint_z \frac{dx}{x-z} <N_0^{ab}(y) \omega_\alpha(x)> \lambda^\beta(z) \nonumber\\
&=& \frac{1}{2} (\Gamma^{cd})^\alpha \ _\beta 
\frac{1}{2} \frac{1}{z-y} (\Gamma^{ab})^\sigma \ _\rho
(\delta^\rho_\alpha - K_\alpha \ ^\rho(y)) 
\Big[ \frac{1}{y-z} (\delta^\beta_\sigma - K_\sigma \ ^\beta(z)) 
+ (\omega_\sigma(y) \lambda^\beta(z)) \Big] \nonumber\\
&-& \frac{1}{4} \frac{1}{y-z} ( \omega \Gamma^{ab} \Gamma^{cd} \lambda )(z) \nonumber\\
&=& - \frac{4}{(y-z)^2} \eta^{d[a} \eta^{b]c} \nonumber\\
&+& \frac{1}{8} \frac{1}{(y-z)^2} \Big[ (Y \Gamma_e \Gamma^{ab} \Gamma^{cd} \Gamma^e \lambda)(z)
+ (Y \Gamma_e \Gamma^{cd} \Gamma^{ab} \Gamma^e \lambda)(y) \Big] \nonumber\\
&-& \frac{1}{16} \frac{1}{(y-z)^2} (Y(y) \Gamma^e \Gamma^{ab} \Gamma_f \lambda(z)) \cdot
(Y(z) \Gamma^f \Gamma^{cd} \Gamma_e \lambda(y))  \nonumber\\
&-& \frac{1}{4} \frac{1}{y-z} ( \omega \Gamma^{ab} \Gamma^{cd} \lambda )(z).
\label{3.18}
\end{eqnarray}
And similarly, 
\begin{eqnarray}
&{}& I_2 \equiv \frac{1}{2} (\Gamma^{cd})^\alpha \ _\beta 
\oint_z \frac{dx}{x-z} \omega_\alpha(x) <N_0^{ab}(y) \lambda^\beta(z)> \nonumber\\
&=& \frac{1}{2} (\Gamma^{cd})^\alpha \ _\beta (\Gamma^{ab})^\beta \ _\rho
\oint_z \frac{dx}{x-z} \omega_\alpha(x) \frac{1}{2} \frac{1}{y-z}  
(\Gamma^{ab} \lambda)^\beta(z) \nonumber\\
&=& \frac{1}{4} (\Gamma^{cd} \Gamma^{ab})^\alpha \ _\rho
\oint_z \frac{dx}{x-z} \frac{1}{y-z} \Big[ \frac{1}{x-z} (\delta^\rho_\alpha - K_\alpha \ ^\rho(z))
+ ( \omega_\alpha(x) \lambda^\rho(z)) \Big] \nonumber\\
&=& \frac{1}{4} \frac{1}{y-z} (\omega \Gamma^{cd} \Gamma^{ab} \lambda).
\label{3.19}
\end{eqnarray}
Thus, we obtain
\begin{eqnarray}
&{}& <N_0^{ab}(y) N_0^{cd}(z)> \nonumber\\
&=& - \frac{4}{(y-z)^2} \eta^{d[a} \eta^{b]c} 
+ \frac{1}{8}\frac{1}{(y-z)^2} \Big[(Y \Gamma_e \Gamma^{ab} \Gamma^{cd}
\Gamma^e \lambda)(z) + (Y \Gamma_e \Gamma^{cd} \Gamma^{ab} \Gamma^e
 \lambda)(y) \Big] \nonumber\\
&-& \frac{1}{16} \frac{1}{(y-z)^2} (Y(y)\Gamma^e \Gamma^{ab} \Gamma_f 
\lambda(z)) \cdot (Y(z)\Gamma^f \Gamma^{cd} \Gamma_e \lambda(y)) \nonumber\\
&-& \frac{1}{4}\frac{1}{y-z} \Big[(\omega \Gamma^{ab} \Gamma^{cd} 
\lambda) - (\omega \Gamma^{cd} \Gamma^{ab} \lambda)(z) \Big].
\label{3.20}
\end{eqnarray}
where note that the final result is symmetric under the simultaneous exchange 
of $(ab) \leftrightarrow (cd)$ and $y \leftrightarrow z$ as desired. 
Then, we can read off the following double and single pole terms from (\ref{3.20}):
\begin{eqnarray}
<N_0^{ab}(y) N_0^{cd}(z)>|_{double \ pole} 
= \frac{1}{(y-z)^2} \Big[ -2 \eta^{d[a} \eta^{b]c} + (Y \Gamma^{abcd} \lambda)
- \frac{3}{4} (Y \Gamma^{ab} \lambda) ( Y \Gamma^{cd} \lambda) \Big](\xi),
\label{3.21}
\end{eqnarray}
and
\begin{eqnarray}
&{}& <N_0^{ab}(y) N_0^{cd}(z)>|_{single \ pole} \nonumber\\
&=& \frac{1}{y-z} \Big[ - (\eta^{a[c} N_0^{d]b} - \eta^{b[c} N_0^{d]a}) 
+ \frac{1}{2} (\eta^{a[c} \partial (Y \Gamma^{d]b}\lambda) 
- \eta^{b[c} \partial (Y \Gamma^{d]a}\lambda)) \nonumber\\
&-& \frac{1}{8} ( (Y \Gamma^{ab} \lambda) \partial (Y \Gamma^{cd} 
\lambda) - (Y \Gamma^{cd} \lambda) \partial (Y \Gamma^{ab} \lambda)) \Big],
\label{3.22}
\end{eqnarray}
where the residue of double pole is evaluated in $\xi = \frac{1}{2}(y+z)$
to make manifest the symmetry under the exchange of $(ab) \leftrightarrow (cd)$
and $y \leftrightarrow z$.
Along the similar line of calculations to the above, we can evaluate the following
two contractions:
\begin{eqnarray}
&{}& \frac{1}{4} \Big( <N_0^{ab}(y) ( \partial Y \Gamma^{cd} \lambda(z)>
+ <N_0^{cd}(z) ( \partial Y \Gamma^{ab} \lambda(y)> \Big) \nonumber\\
&=& - \frac{1}{4} \frac{1}{(y-z)^2} (Y \Gamma^{ab} \lambda) \cdot
(Y \Gamma^{cd} \lambda)(\xi) \nonumber\\
&+& \frac{1}{4} \frac{1}{y-z} \Big[ - \partial (Y \Gamma^{ab} \lambda) \cdot
(Y \Gamma^{cd} \lambda) - \eta^{a[c} (\partial Y \Gamma^{d]b}\lambda) 
+ \eta^{b[c} (\partial Y \Gamma^{d]a}\lambda) \Big],
\label{3.23}
\end{eqnarray}
and
\begin{eqnarray}
&-& \Big( <N_0^{ab}(y) \partial (Y \Gamma^{cd} \lambda)(z)>
+ <N_0^{cd}(z) \partial (Y \Gamma^{ab} \lambda)(y)> \Big) \nonumber\\
&=& - \frac{1}{(y-z)^2} \Big[ \eta^{d[a} \eta^{b]c} + (Y \Gamma^{abcd} \lambda)
-  (Y \Gamma^{ab} \lambda) ( Y \Gamma^{cd} \lambda) \Big](\xi)
\nonumber\\
&+& \frac{1}{2} \frac{1}{y-z} \Big[ \partial \Big( (Y \Gamma^{ab} \lambda)
(Y \Gamma^{cd} \lambda) \Big) - \partial ( Y \Gamma^{cd} \Gamma^{ab} \lambda) \Big].
\label{3.24}
\end{eqnarray}
The two last spurious terms in the double pole (\ref{3.21}) cancel if we 
choose for the definition of $N^{ab} $ the linear combination given in 
(\ref{3.8}). With this choice the spurious terms in the single pole also cancel
automatically, so one recovers the desired Lorentz algebra with anomaly $-3$ 
given in (\ref {3.6}). As a remark, in order to respect the symmetry 
under the exchange of $(ab)\leftrightarrow (cd)$ and $y \leftrightarrow z$
in the OPE's, we have used the midpoint $\xi$ parametrization in the above, 
but 
instead we can make use of the conventional $z$ parametrization. In fact,
we have checked that we can derive the OPE (\ref {3.6}), but to do so,
we need a nontrivial identity \cite{Luca}
\begin{eqnarray}
\partial (Y \Gamma^{abcd} \lambda)
&=& \partial \Big( (Y \Gamma^{ab} \lambda)(Y \Gamma^{cd} \lambda) \Big)
- (Y \Gamma^{b[c} \lambda) \partial (Y \Gamma^{d]a} \lambda)
+ (Y \Gamma^{a[c} \lambda) \partial (Y \Gamma^{d]b} \lambda) \nonumber\\
&=& 4 (Y \Gamma^{[ab} \lambda) \partial (Y \Gamma^{cd]} \lambda).
\label{3.25}
\end{eqnarray}

Next let us consider the $J$-$N^{ab}$ OPE. We can easily calculate
\begin{eqnarray}
<J_0 (y) N^{ab}_0 (z)> = - \frac{3}{2} \frac{1}{(y-z)^{2}}(Y\Gamma^{ab}
\lambda)(z),
\label{3.26}
\end{eqnarray}
\begin{eqnarray}
<J_0 (y) (\partial Y \Gamma^{ab}\lambda)(z)> = - \frac{1}{(y-z)^{2}}(Y\Gamma^{ab}
\lambda)(z),
\label{3.27}
\end{eqnarray}
\begin{eqnarray}
<J_0 (y)(Y \Gamma^{ab}\partial \lambda)(z)> = {\frac{1}{(y-z)^{2}}}(Y\Gamma^{ab}
\lambda)(z),
\label{3.28}
\end{eqnarray}
\begin{eqnarray}
<(\partial Y \lambda)(y) N^{ab}_0 (z)> = - \frac{1}{2} \frac{1}{(y-z)^{2}}
(Y \Gamma^{ab} \lambda)(z).
\label{3.29}
\end{eqnarray}
Then, given Eqs. (\ref{3.8}) and (\ref{3.9}), one can see that the spurious 
double pole in the OPE between $J$ and $N^{ab}$ drops out so that
(\ref{3.5}) is obtained.

To compute the OPE between $T$ and $N^{ab} $ we need the 
following OPE's:
\begin{eqnarray}
<T^{(\omega\lambda)}_0 (y) N^{ab}_0 (z)> = \frac{3}{2}\frac{1}{(y-z)^{3}}
(Y\Gamma^{ab}\lambda)(z) + \frac{1}{2} \frac{1}{(y-z)^{2}}(\omega\Gamma^{ab}
\lambda)(z) + \frac{1}{2}\frac{1}{y-z}\partial(\omega\Gamma^{ab}\lambda)(z),  
\label{3.30}
\end{eqnarray}
\begin{eqnarray}
<T^{(\omega\lambda)}_0 (y) (\partial Y \Gamma^{ab}\lambda)(z)> = 
\frac{1}{(y-z)^2}(\partial Y \Gamma^{ab}\lambda)(z) + \frac{1}{y-z}
\partial (\partial Y \Gamma^{ab}\lambda)(z),
\label{3.31}
\end{eqnarray}
\begin{eqnarray} 
<T^{(\omega\lambda)}_0 (y) ( Y \Gamma^{ab}\partial\lambda)(z)> = 
\frac{1}{(y-z)^2}( Y \Gamma^{ab}\partial\lambda)(z) + \frac{1}{y-z}
\partial ( Y \Gamma^{ab}\partial\lambda)(z),
\label{3.32}
\end{eqnarray}
\begin{eqnarray}
<\partial(Y\partial\lambda)(y)N^{ab}_0 (z)> = - \frac{1}{(y-z)^3}
(Y\Gamma^{ab} \lambda)(z).
\label{3.33}
\end{eqnarray}
As before, given (\ref{3.7}), (\ref{3.8}) and (\ref{3.11}), the spurious term 
vanishes, thereby reproducing the correct OPE between $T$ and $N^{ab}$ 
as given in (\ref{3.3}).

The $T$-$T$ OPE (\ref{3.1}) can be evaluated by using 
\begin{eqnarray}
<T^{(\omega\lambda)}_0 (y) T^{(\omega\lambda)}_0 (z)> &=& 
\frac{11}{(y-z)^4} + \frac{2}{(y-z)^2} T^{(\omega\lambda)}_0 (\xi),
\label{3.34}
\end{eqnarray}
\begin{eqnarray}
<T^{(\omega\lambda)}_0 (y)\partial(Y\partial\lambda)(z)> +    
 <\partial(Y\partial\lambda)(y)T^{(\omega\lambda)}_0 (z)> =
\frac{2}{(y-z)^{2}}\partial(Y\partial\lambda)(\xi).
\label{3.35}
\end{eqnarray}  

Similarly, the $T$-$J$ OPE (\ref{3.2}) is obtained from
\begin{eqnarray}
<T^{(\omega\lambda)}_0 (y)J_0 (z)>= \frac{11}{(y-z)^{3}} + \frac{1}{(y-z)^2}J_0 (z) 
+ \frac{1}{y-z}\partial J_0 (z),
\label{3.36}
\end{eqnarray}  
\begin{eqnarray}
<\partial(Y \partial\lambda)(y)J_0 (z)>= -\frac{2}{(y-z)^{3}},
\label{3.37}
\end{eqnarray}  
\begin{eqnarray}
<T^{(\omega\lambda)}_0 (y)(\partial Y\lambda)(z)>= \frac{1}{(y-z)^{2}}
(\partial Y\lambda)(z) + \frac{1}{y-z} \partial(\partial Y \lambda)(z),
\label{3.38}
\end{eqnarray}  
and the $J$-$J$ OPE (\ref{3.4}) follows from
\begin{eqnarray}
<J_0 (y)J_0 (z)> = - \frac{11}{(y-z)^{2}},
\label{3.39}
\end{eqnarray}  
\begin{eqnarray}
<J_0 (y)(\partial Y \lambda)(z)> + <(\partial Y \lambda)(y) J_0 (z)> =
- \frac{2}{(y-z)^2}.
\label{3.40}
\end{eqnarray}  
In this way, we have succeeded in deriving the whole OPE's (\ref{3.1})-(\ref{3.6})
when we postulate the expressions for $T$, $N^{ab}$ and $J$ as in 
Eqs. (\ref{3.7})-(\ref{3.9}). Moreover, it is straightforward to construct
the unintegrated and integrated vertex operators by adding the ASPC operators
and calculate the amplitudes within the framework of Y-formalism.
Finally, we should stress that even if we have the OPE between $\omega$
and $\lambda$ which is not completely free because of the presence of the 
projector $K$
we can construct a self-consistent conformal field theory for GS superstrings.

\section{The b-field}

In this section, following \cite{Oda2} we shall prove the equivalence of 
cohomology between the field $b$, as given in (\ref{2.22}), times $Z_B$ and the 
picture raised $b$ field $b_B$ as given in \cite{Ber6} and we shall give the
full expression of $b_B$. The style of this section will be quite different 
from that of the previous section in that we work at a classical level,
thereby ignoring normal-ordering contributions and corrections from double
contractions. Accordingly, in most of parts in this 
section we will come back to the notations of section 2 for the quantites 
defined in (\ref{2.8}), (\ref{2.9}) and (\ref{2.21}), dropping the distinction 
between $N^{ab}$, $J$, $T$ and $N^{ab}_0$, $J_0$, $T_0$.   

The recipe of Berkovits \cite{Ber6} to compute $g$-loops amplitudes is based 
on three ingredients:
\par i) the Lorentz invariant measure for pure spinor ghosts and 
their momenta. 
\par ii) the picture changing operators (PCO's) to cancel the zero modes of 
the ghosts, i.e., $11$ insertions of the PCO's, $Y_C$ to absorb the $11$ zero 
modes of $\lambda$ and $11g$ insertions of the PCO's  $Z \equiv (Z_{B},Z_{J})$
 to absorb the $11g$ zero modes of $\omega$.
\par iii) $(3g - 3)$ insertions ($1$ at  $g=1$) of $\int\mu b$ where $\mu$ is 
the Beltrami parameter and $b$ is the $b$ field, in order to take account of 
the moduli of the genus $g$ Riemann surface.

The PCO's are BRST-closed and their derivative is BRST-exact. The latter 
requirement ensures that the amplitudes are independent of the position of 
insertions of the PCO's. According to \cite{Ber6}, the PCO's are given by
\begin{eqnarray}
Y_C = (C\theta) \delta(C\lambda),
\label{4.1}
\end{eqnarray}
\begin{eqnarray}
Z_B = \frac{1}{2} B_{ab} (\lambda \Gamma^{ab} d)\delta (B_{cd}N^{cd}),
\label{4.2}
\end{eqnarray}   
\begin{eqnarray}
Z_J = (\lambda d)\delta(J),
\label{4.3}
\end{eqnarray}
where $C_{\alpha}$ and $B^{ab}$ are constants to be integrated later. The 
unavoidable breaking of the super-Poincar\'e invariance
implied by the presence of these constants is common to all theories with 
zero modes which belong to non-scalar Lorentz representations. However, the 
amplitudes 
still remain to be covariant since, as shown in \cite{Ber6}, the 
super-Poincar\'e variation of these PCO's is BRST-exact.  

As for the $b$ field, it is defined in (\ref{2.22}) and its main property is 
(\ref{2.20})
This $b$ field has two disturbing features due to its dependence on $Y_\alpha$:
the first feature is that it is not covariant and the second one is that it
becomes singular for the configurations where $(v \lambda) = 0$. The first 
feature is perhaps not so dangerous since the variation $\delta_L b $ of $b$ under global 
Lorentz 
transformations is BRST-exact. Indeed, since $\{ Q, b \} = T$ and $\delta_L Q =
\delta_L T = 0$, this implies that
\begin{eqnarray}
\delta_L b = [ Q, \Omega ],
\end{eqnarray}
where $\Omega$ is some local function of the ghost number $0$. One has indeed
\begin{eqnarray}
\Omega = {\frac{1}{2}} (\tilde \omega \Pi^m\Gamma_m \delta Y) + {\frac{1}{8}}
\Lambda_{mn}(Y\Gamma^m d)(Y\Gamma^n d),
\label{4.5}
\end{eqnarray}
where $\delta Y_\alpha = 
{\frac{1}{2}}\Lambda_{mn}(Y\Gamma^{mn})_\beta (\delta^
{\beta}_\alpha - \lambda^\beta Y_\alpha) $ and $\Lambda_{mn}$ are global 
Lorentz parameters.
  
To circumvent these difficulties,
Berkovits \cite{Ber6} replaced $b$ with a picture raised, covariant $b$ field 
denoted as $b_B$, whose BRST variation reads
\begin{eqnarray}
[ Q, b_B ] = T Z_B.
\label{4.6}
\end{eqnarray}
In \cite{Oda2} we have shown that $b$ and $b_B$ are closely related in the 
sense that, if we associate to $b$ a picture raising operator $Z_B$, 
then $b Z_B$ and $b_B$ are cohomologically equivalent
\begin{eqnarray}
b Z_B = b_B + [ Q, X ].
\label{4.7}
\end{eqnarray}
In this section we not only review the result of \cite{Oda2} but also we 
derive, using our formalism, the complete expression of $b_B$. 

Before doing that, we need some more properties of the picture raising 
operators $Z$ 
given in (\ref{4.2}), (\ref{4.3}) and of the field $b$ (and $b_B$).
Two definitions are in order \par
{\it \underline{ Definition I}}: A superfield $X^{\alpha_1 \cdots \alpha_n}$ 
\begin{eqnarray}
X^{\alpha_1 \cdots \alpha_n} =   \sum_{i} h_i^{\alpha_1 \cdots (\alpha_i
\alpha_{i+1}) \cdots \alpha_n}, 
\label{4.8}
\end{eqnarray}
such that $ h_i^{\alpha_1 \cdots (\alpha_i \alpha_{i+1}) \cdots \alpha_n}$ is 
symmetric in the indices $(\alpha_i \alpha_{i+1})$ and satisfies 
\begin{eqnarray}
(\Gamma^a)_{\alpha_i \alpha_{i+1}} h_i^{\alpha_1 \cdots (\alpha_i \alpha_{i+1}) \cdots
\alpha_n} = 0,
\label{4.9}
\end{eqnarray}
is called $\Gamma_1$-{\it traceless}.

{\it \underline{ Definition II}}: A quantity $Y_{\alpha_1 \cdots \alpha_n}$
is called  $\Gamma_5$-{\it traceless} if it vanishes when saturated  with 
$(\Gamma_{a_1 \cdots a_5})^{\alpha_i \alpha_{i+1}}$ between two adjacent 
indices.

It is important to notice that a $\Gamma_1$-traceless superfield vanishes if 
saturated with a $\Gamma_5$-traceless object.

Let us come back to the BRST-closed, picture raising operator $Z_B \equiv 
\lambda^\alpha Z_\alpha$ defined in (\ref{4.2}) which satisfies
\begin{eqnarray}
\{ Q, Z_B \} = 0. 
\label{4.10}
\end{eqnarray}
Then the following recursion equations hold:
\begin{eqnarray}
\{ Q, Z_\alpha \} = \lambda^\beta Z_{\beta\alpha},
\label{4.11}
\end{eqnarray}
\begin{eqnarray}
[ Q, Z_{\beta\alpha} ]  = \lambda^\gamma Z_{\gamma\beta\alpha},
\label{4.12}
\end{eqnarray}
\begin{eqnarray}
\{ Q, Z_{\gamma\beta\alpha} \} = \lambda^\delta Z_{\delta\gamma\beta\alpha} 
+ \partial \lambda^\delta \Upsilon_{\delta\gamma\beta\alpha},
\label{4.13}
\end{eqnarray} 
where the operators $Z_{\beta\alpha}$, $Z_{\gamma\beta\alpha}$, $Z_{\delta\gamma
\beta\alpha}$ and $\Upsilon_{\delta\gamma\beta\alpha}$ are $\Gamma_5$-{\it traceless}.

This result can be easily derived from the expressions of $Z$ in (\ref{4.2})
but also follows from (\ref{4.10}), the pure spinor constraint 
and the fact that $Z$ depends only on $\lambda$, $\omega$, and $d$ so that 
$\partial \lambda$ can appear only at the third step. 
Concretely, the full expression of the operators $Z$'s and $\Upsilon_{
\alpha_1 \cdots \alpha_4}$ in the case of $Z_B$ takes the form \cite{Ber6}
\begin{eqnarray}
Z_\alpha &=& \frac{1}{2} B_{ab} (\Gamma^{ab} d)_\alpha \delta(B_{cd} N^{cd}) \nonumber\\
&\equiv& \frac{1}{2} (B d)_\alpha \delta(B N),
\label{4.14}
\end{eqnarray}       
\begin{eqnarray}
Z_{\beta\alpha} = - \frac{1}{2} (\Gamma^c \Gamma^{ba})_{\beta\alpha} \Pi_c B_{ab} \delta(B N)
- \frac{1}{4} (B d)_\alpha (B d)_\beta \partial \delta(B N),
\label{4.15}
\end{eqnarray}       
\begin{eqnarray}
Z_{\gamma\beta\alpha} &=& - \frac{1}{2} \Big[ (\Gamma^c \Gamma^{ba})_{\beta\alpha} 
(\Gamma_c \partial \theta)_\gamma B_{ab} \delta(B N)
+ \frac{1}{2} (\Gamma^c \Gamma^{ba})_{\beta\alpha} (B d)_\gamma \Pi_c B_{ab} \partial \delta(B N)
\nonumber\\
&+& \frac{1}{2} (\Gamma^c \Gamma^{ba})_{\gamma[\beta} (B d)_{\alpha]} \Pi_c B_{ab} \partial \delta(B N)
+ \frac{1}{4}  (B d)_\alpha (B d)_\beta (B d)_\gamma \partial^2 \delta(B N) \Big],
\label{4.16}
\end{eqnarray}       
\begin{eqnarray}
Z_{\delta\gamma\beta\alpha} &=& - \frac{1}{4} \Big[ \Big( (\Gamma^c \Gamma^{ba})_{\beta\alpha}
(B d)_{[\delta} (\Gamma_c \partial \theta)_{\gamma]} 
- (\Gamma^c \Gamma^{ba})_{\gamma[\beta} (B d)_{\alpha]} (\Gamma_c \partial \theta)_{\delta} \Big)
B_{ab} \partial \delta(B N) \nonumber\\
&-& \Big(  (\Gamma^f \Gamma^{ed})_{\delta[\gamma} (\Gamma^c \Gamma^{ba})_{\beta]\alpha}
+  (\Gamma^f \Gamma^{ed})_{\delta\alpha} (\Gamma^c \Gamma^{ba})_{\gamma\beta} \Big)
\Pi_c B_{ab} \Pi_f B_{de} \partial \delta(B N) \nonumber\\
&-& \frac{1}{2} \Big( (\Gamma^c \Gamma^{ba})_{\beta\alpha} (B d)_\gamma (B d)_\delta
+ (\Gamma^c \Gamma^{ba})_{\gamma[\beta} (B d)_{\alpha]} (B d)_\delta  \nonumber\\
&+& \frac{1}{2} (\Gamma^c \Gamma^{ba})_{\delta[\alpha} (B d)_\beta (B d)_{\gamma]} \Big)
\Pi_c B_{ab} \partial^2 \delta(B N)  \nonumber\\
&-& \frac{1}{4}  (B d)_\alpha (B d)_\beta (B d)_\gamma (B d)_\delta \partial^3 \delta(B N) \Big],
\label{4.17}
\end{eqnarray}       
\begin{eqnarray}
\Upsilon_{\delta\gamma\beta\alpha} = - \frac{1}{2} (\Gamma_c)_{\delta\gamma}
(\Gamma^c \Gamma^{ba})_{\beta\alpha} B_{ab} \delta(B N).
\label{4.18}
\end{eqnarray}       
In (\ref{4.15})-(\ref{4.17}), the symbol $\partial$ applied to $\delta(BN)$
 means the derivative of $\delta(BN)$ with respect to its argument. 

As a further preliminary, it is convenient to record the following identity
\begin{eqnarray}
-{\frac{1}{8}}(B\Gamma^{ab}A)(\Gamma_{ab}C)^\alpha 
-{\frac{1}{4}}(BA)C^\alpha =
B_\beta A^\alpha C^\beta - {\frac{1}{2}}(\Gamma^a B)^\alpha (A\Gamma_a C), 
\label{4.19}
\end{eqnarray}
where $A^\alpha$, $B_\beta$, $C^\gamma$ are generic spinor-like quantities (a 
similar identity holds for $A_\alpha$, $B^\beta$, $C_\gamma$). 
Eq. (\ref{4.19}) can be proved with the help of the obvious identity 
\begin{eqnarray}
A^\alpha B_\beta = {\frac{1}{16}} \delta^{\alpha}_{\beta} (B A) 
- \frac{1}{16} (\Gamma_{ab})^{\alpha} \ _{\beta} {\frac{1}{2}}(B\Gamma^{ab}A) + 
{\frac{1}{384}}(\Gamma^{abcd})^\alpha\ _\beta (B\Gamma_{abcd}A).
\nonumber
\end{eqnarray}
In particular, with  $ A^\alpha = \lambda^\alpha$, 
$B_\beta = \omega_\beta$ and $C^\gamma = \partial\theta^\gamma $, the identity 
(\ref{4.19}) becomes
\begin{eqnarray}
\lambda^\alpha (\omega \partial \theta) - {\frac{1}{2}}(\Gamma^a\omega)^\alpha
(\lambda \Gamma_a \partial \theta) = - {\frac{1}{4}}N_{ab}(\Gamma^{ab}\partial
 \theta)^\alpha - {\frac{1}{4}} J \partial \theta ^\alpha.
\label{4.20}
\end{eqnarray}
Now let us turn our attention to the $b$ field.
Making use of $(Y\lambda)=1$, $b$ in (\ref{2.22}) can be rewritten as
\begin{eqnarray}
b = {\frac{1}{2}}Y_\alpha [(\Pi^a \Gamma_a d)^\alpha + \lambda^\alpha (\omega
\partial\theta) - {\frac{1}{2}} (\Gamma^a\omega)^\alpha (\lambda \Gamma_a 
\partial \theta) ],
\label{4.21}
\end{eqnarray}
which, using the identity (\ref{4.20}), becomes 
\begin{eqnarray}
b = Y_\alpha G^\alpha ,
\label{4.22}
\end{eqnarray}
where
\begin{eqnarray}
G \ ^{\alpha} = \frac{1}{2}\Pi_a (\Gamma^a d)^\alpha - \frac{1}{4} 
N^{ab} (\Gamma_{ab}\partial \theta)^{\alpha} - \frac{1}{4} J \partial \theta^
\alpha.
\label{4.23}
\end{eqnarray}

The covariant field $G^\alpha$ has first appeared in \cite{Ber4} and its BRST
variation, at least at classical level, is
 \begin{eqnarray}
\{ Q, G^\alpha \} = \lambda^\alpha T. 
\label{4.24}
\end{eqnarray}
Eq. (\ref{4.24}) also follows  from $\{ Q, b \} = T \equiv 
(Y_\alpha \lambda^\alpha) T$, (\ref{4.22}) and the fact that $G^\alpha$ and 
$\lambda^\alpha T$ are covariant.
Note that Eq. (\ref{4.24}) is the basis of the Berkovits' construction of 
$b_B$.
Indeed in \cite{Ber6} it was proved that Eq. (\ref{4.24}) implies the existence
 of fields $H^{\alpha\beta}$, $K^{\alpha\beta\gamma}$, $L^{\alpha\beta\gamma
\delta}$ (and $S^{\alpha\beta\gamma}$), defined modulo $\Gamma_1$-traceless 
terms that satisfy 
the recursion equations (in a sense dual to (\ref{4.11})-(\ref{4.13}))
\begin{eqnarray}
[Q, H^{\alpha\beta}] = \lambda^\alpha G^\beta + \cdots,
\label{4.25}
\end{eqnarray}
\begin{eqnarray}
\{ Q, K^{\alpha\beta\gamma} \} = \lambda^\alpha H^{\beta\gamma} +
\cdots,
\label{4.26}
\end{eqnarray}
\begin{eqnarray}
[Q, L^{\alpha\beta\gamma\delta}] = \lambda^\alpha K^{\beta\gamma \delta} 
+ \cdots,
\label{4.27}
\end{eqnarray}
where henceforth the dots denote $\Gamma_1$-traceless terms.
The chain of equations finishes at the level of (\ref{4.27}) since, for 
dimensional 
reasons, $\lambda^\alpha L^{\beta\gamma\delta\epsilon}$ vanishes modulo
$\Gamma_1$-traceless terms so that 
\begin{eqnarray}
L^{\alpha\beta\gamma\delta} = \lambda^\alpha S^{\beta\gamma\delta} + \cdots,
\label{4.28}
\end{eqnarray}
and
\begin{eqnarray}
[ Q, S^{\alpha\beta\gamma} ] = K^{\alpha\beta\gamma} + \lambda^{\alpha}T^
{\beta \gamma} + \cdots,
\label{4.29}
\end{eqnarray}
for a suitable field $T^{\beta\gamma}$.
Then, according to \cite{Ber6}, $b_B$ is 
\begin{eqnarray}
b_B = b_1 + b_2 + b_3 + b_4^{(a)} + b_4^{(b)}, 
\label{4.30}
\end{eqnarray}
where 
\begin{eqnarray}
b_1 &=& G^\alpha Z_\alpha, \nonumber\\
b_2 &=& H^{\alpha\beta} Z_{\alpha\beta}, \nonumber\\
b_3 &=& - K^{\alpha\beta\gamma} Z_{\alpha\beta\gamma}, \nonumber\\ 	
b_4^{(a)} &=& - L^{\alpha\beta\gamma\delta}Z_{\alpha\beta\gamma\delta}, 
\nonumber\\
b_4^{(b)} &=& - S^{\alpha\beta\gamma}\partial\lambda^\delta 
\Upsilon_{\delta\alpha\beta\gamma}.
\label{4.31}
\end{eqnarray}
Whereas the expressions of $G^\alpha$ and $H^{\alpha\beta}$ (recall that 
$H^{\alpha\beta}$ 
etc. are defined modulo $\Gamma_1$-traceless operators) are given in \cite
{Ber6},
$K^{\alpha\beta\gamma}$ and $S^{\alpha\beta\gamma}$ (and $L^{\alpha\beta
\gamma\delta}$) are not 
computed explicitly in \cite{Ber6, Oda2}. We will present these expressions 
shortly by using Y-formalism at hand.

A simple way to show (\ref{4.30}) and at the same time the equivalence of
cohomology between $b Z_B$ and $b_B$ is described as follows \cite{Oda2}: 
starting from
\begin{eqnarray}
b Z_B \equiv (Y_\alpha G^\alpha)(\lambda^\beta Z_\beta),
\label{4.32}
\end{eqnarray}
and adding and subtracting the term $b_1 = (Y_{\alpha}\lambda^{\alpha})
G^\beta Z_\beta$, one obtains
\begin{eqnarray}
b Z_B &=& b_1 + Y_\alpha (G^\alpha \lambda^\beta - \lambda^\alpha G^\beta)Z_
\beta \nonumber \\ 
&=& b_1  - [ Q, Y_\alpha (H^{\alpha\beta} - H^{\beta\alpha}) ] Z_\beta 
\nonumber\\ 
&=& b_1 + Y_\alpha (H^{\alpha\beta} - H^{\beta\alpha}) \{ Q, Z_\beta \} + 
\{ Q, X_1 \} \nonumber \\
&=& b_1 + Y_\alpha (H^{\alpha\beta} -H^{\beta\alpha})\lambda^\gamma 
Z_{\gamma\beta} + \{ Q, X_1 \}, 
\label{4.33}
\end{eqnarray}                
where $X_1 = - Y_\alpha(H^{\alpha\beta} - H^{\beta\alpha}) Z_\beta$. The second step follows 
by noting that in $\lambda^\alpha G^\beta - \lambda^\beta 
G^\alpha$ the $\Gamma_1$-traceless  terms drop out so that  one can use
(\ref{4.25}) without the dots and the last step follows from (\ref{4.11}).
This procedure can be repeated again and again by adding and subtracting step 
by step the terms 
\begin{eqnarray}
b_2 &=& (Y_\alpha \lambda^\alpha) H^{\beta\gamma} Z_{\beta\gamma}, \nonumber\\    
b_3 &=& - (Y_\alpha \lambda^\alpha) K^{\beta\gamma\delta} Z_{\beta\gamma\delta},
\nonumber\\    
b_4^{(a)} &=& - (Y_\alpha \lambda^\alpha) L^{\beta\gamma\delta\epsilon} Z_{\beta
\gamma\delta\epsilon},
 \label{4.34}
\end{eqnarray}
where we have used the fact that whenever we meet the linear combinations of 
$\lambda H$,
$\lambda K$ and $\lambda L $ which arise at each step, the 
$\Gamma_1$-traceless terms drop out. Eventually we end with 
\begin{eqnarray}
 b Z_B = b_1 + b_2 + b_3 + b_4^{(a)} + \lbrace Q, X_B \rbrace
- Y_\alpha \lambda^\alpha Y_\eta L^{\eta\delta\gamma\beta}\partial \lambda^
\epsilon \Upsilon_{\epsilon\delta\gamma\beta},
\label{4.35}
\end{eqnarray}
where $X_B$ is defined as
\begin{eqnarray}
X_B &=& - Y_\alpha (H^{\alpha\beta} - H^{\beta\alpha}) Z_\beta \nonumber\\    
&+& Y_\alpha (K^{\gamma\alpha\beta} - K^{\gamma\beta\alpha} - K^{\alpha\gamma\beta})
Z_{\gamma\beta} \nonumber\\
&+& Y_\alpha (L^{\delta\gamma\alpha\beta} - L^{\delta\gamma\beta\alpha} 
- L^{\delta\alpha\gamma\beta} + L^{\alpha\delta\gamma\beta})
Z_{\delta\gamma\beta}.
 \label{4.36}
\end{eqnarray}
Let us notice, using (\ref{2.12}) and (\ref{4.28}),  that the last term in 
(\ref{4.35}) is nothing but $ b_4^{(b)}$.
Thus, we arrive at Eq. (\ref{4.7}) from Eq. (\ref{4.30}). Moreover, since 
$\{ Q, b \} = T$ and therefore  $[ Q, b Z_B ] = T Z_B$, we also obtain Eq. (\ref{4.6}).   
  
Now we are ready to give the full expressions of the fields
 $H^{\alpha\beta}$, $K^{\alpha\beta\gamma}$, $L^{\alpha\beta\gamma\delta}$ and 
$S^{\alpha\beta\gamma}$ that enter in the construction of $b_B$. 
It is of interest that the formalism at hand
allows us to obtain such complicated expressions with a reasonable amount 
of effort.
They are obtained by starting with the ansatz that the 
fields $H$, $K$, $L$, and $S$ are constructed out of only the building 
blocks $\lambda$, $\Gamma^{ab}\lambda$, $\Gamma^a \tilde\omega$ and 
$\Gamma^a d$
(as well as $\Pi^a$ in $H$), writing their more general expressions in 
terms of these blocks, and moreover imposing the condition that in the fields
 $H$, $K$, any dependence on $Y_\alpha$ which is hidden in $\tilde
\omega$, 
should be absent. These fields are then  determined by requiring 
that they satisfy Eqs. (\ref{4.25})-(\ref{4.29}). 
 A solution for $H^{\alpha\beta}$, $K^{\alpha\beta\gamma}$, $S^{\alpha\beta
\gamma}$ and $L^{\alpha\beta\gamma\delta}$ is
\begin{eqnarray}
H^{\alpha\beta} = - {1\over 16}(\Gamma^a d)^\alpha (\Gamma_a d)^\beta 
- {1\over 2}\lambda^\alpha \Pi^a (\Gamma_a \tilde\omega)^\beta   
+{1\over 16} \Big[ \Pi^a (\Gamma_b \Gamma_a \lambda)^\alpha
(\Gamma^b \tilde\omega)^\beta - (\alpha \leftrightarrow \beta)\Big],
\label{4.37}
\end{eqnarray}
\begin{eqnarray}
K^{\alpha\beta\gamma} &=& {1\over 16}\lambda^\alpha
(\Gamma^a \tilde\omega)^\beta (\Gamma_a d)^\gamma 
+ {1\over 32} \Big[(\tilde\omega\Gamma_a)^\alpha \lambda^\beta(\Gamma^a d)^\gamma 
+ (\alpha \leftrightarrow \gamma)\Big]   \nonumber\\
&+& {1\over 96} \Big[(\tilde\omega\Gamma^a)^\alpha (\Gamma_{ab}\lambda)^\beta 
(\Gamma^b d)^\gamma - (\alpha\leftrightarrow \gamma)\Big],
\label{4.38}
\end{eqnarray}
\begin{eqnarray}
S^{\alpha\beta\gamma} = - {1\over 32} (\tilde\omega\Gamma_a)^\alpha\lambda^\beta 
(\Gamma^a \tilde\omega)^\gamma - {1\over 96}(\tilde\omega \Gamma^a)^\alpha
(\Gamma_{ab}\lambda)^\beta (\Gamma^b \tilde\omega)^\gamma,  
\label{4.39}
\end{eqnarray}
so that, from (\ref{4.28}) 
\begin{eqnarray}
L^{\alpha\beta\gamma\delta} &=& - {1\over 32} \lambda^\alpha (\tilde\omega
\Gamma^a)^\beta
\Big[ \lambda^\gamma (\Gamma_a \tilde\omega)^\delta + {1\over 3}(\Gamma_{ab}
\lambda)^\gamma
(\Gamma^b \tilde\omega)^\delta \Big]. 
\label{4.40}
\end{eqnarray}
Recall that the field $H$, $K$ 
 are not only Lorentz covariant but also invariant under the 
gauge symmetry (\ref{2.7}) (since $\tilde\omega$ is invariant). 
Therefore they should depend on $\omega$ only through $N^{ab}$ and $J$. 

Indeed, using Eq. (\ref{4.21}) one obtains
\begin{eqnarray}
H^{\alpha\beta} = {1\over 16} \Gamma_a^{\alpha\beta} 
\Big( N^{ab} \Pi_b - {1\over 2} J \Pi^a \Big)
+ {1\over 384} \Gamma_{abc}^{\alpha\beta} 
\Big( d \Gamma^{abc} d + 24 N^{ab} \Pi^c \Big),
\label{4.41}
\end{eqnarray}
which precisely coincides with the expression given in \cite{Ber6}, and        

\begin{eqnarray}
K^{\alpha\beta\gamma} &=& - {1\over 48} \Gamma_a^{\alpha\beta} (\Gamma_b d)^\gamma N^{ab}
- {1\over 192} \Gamma_{abc}^{\alpha\beta} (\Gamma^a d)^\gamma N^{bc} \nonumber\\
&+& {1\over 192} \Gamma_a^{\gamma\beta} \Big[ (\Gamma_b d)^\alpha N^{ab}
+ {3 \over 2} (\Gamma^a d)^\alpha J \Big]
+ {1\over 192} \Gamma_{abc}^{\gamma\beta} (\Gamma^a d)^\alpha N^{bc}.
\label{4.42}
\end{eqnarray}

Furthermore, it is easy to derive from (\ref{4.39}) and (\ref{4.18}) the 
complete expression of $b^{(b)}_4$ whose result is given by
\begin{eqnarray}
b^{(b)}_4 = B_{ab} \Big[ -  T^{\omega\lambda} N^{ab} - {1\over 4} J 
\partial N^{ab}
+ {1\over 4} N^{ab} \partial J + {1\over 2} N^a \ _c \partial N^{bc} \Big] 
\delta(B N).
\label{4.43}
\end{eqnarray}
This equation clearly shows that the Y-dependence of $S^{\alpha\beta\gamma}$ 
disappears in $b^{(b)}_4$.

Finally, let us consider $L^{\alpha\beta\gamma\delta}$ in detail.
We shall show that $L^{\alpha\beta\gamma\delta}$ has a residual dependence on
$Y_\alpha$,
but it turns out that the $Y_\alpha$-dependent terms are $\Gamma_1$-traceless 
and therefore do not contribute to $b_B$.
Indeed, we can rewrite $L^{\alpha\beta\gamma\delta}$ in (\ref{4.40}) as 
follows:
\begin{eqnarray}
L^{\alpha\beta\gamma\delta} = - \frac{1}{32} \Big[ A_{a}^{\alpha\beta} - (\tilde\omega
\Gamma_a) ^\alpha \lambda^\beta\Big] B^{a\gamma\delta},
\label{4.44}
\end{eqnarray}
where 
\begin{eqnarray} 
A_{a}^{\alpha\beta} &=& \lambda^\alpha (\tilde\omega\Gamma_a)^\beta  +  
(\tilde\omega\Gamma_a)^\alpha \lambda^\beta \nonumber \\ 
&=& {\frac{1}{4}} (\Gamma_c)^{\alpha\beta}N^{ac} +{\frac{1}{8}}(\Gamma_a)^{\alpha\beta}J
+ \cdots,
\label{4.45}
\end{eqnarray} 
and
\begin{eqnarray}
B_{a}^{\gamma\delta} &=& \frac{4}{3} \lambda^\gamma(\Gamma_a\tilde \omega)^\delta -
\frac{1}{3} (\Gamma_b \Gamma_a\lambda)^\gamma (\Gamma^b \tilde\omega)^\delta
\nonumber \\ 
&=&  \frac{1}{4} (\Gamma_a)^{\gamma\delta} J - \frac{1}{6} (\Gamma_{ab_1b_2})^{\gamma\delta}
N^{b_1b_2} + \frac{1}{6} (\Gamma_b)^{\gamma\delta} N^b \ _a + \cdots, 
\label{4.46}
\end{eqnarray}
depend on only $N^{ab}$ and $J$ modulo $\Gamma_1$-traceless terms.  However
\begin{eqnarray}
(\tilde\omega\Gamma_a)^\alpha \lambda^\beta = \frac{1}{16} \Big[ (\Gamma_a)^
{\alpha\beta}J + (\Gamma_a\Gamma_{bc})^{\alpha\beta}N^{bc} + \frac{1}{24}
(\Gamma_a \Gamma_{b_1 \cdots b_4})^{\alpha\beta} (\tilde\omega\Gamma^{ b_1 \cdots b_4} \lambda)
\Big],
\label {4.47}
\end{eqnarray}
keeps a dependence on $Y_\alpha$,  which will be denoted as 
$(\tilde\omega\Gamma_a)^\alpha \lambda^\beta|_Y$, given by
\begin{eqnarray}
(\tilde\omega\Gamma_a)^\alpha \lambda^\beta|_Y 
&=& - \frac{1}{2}(\omega\Gamma^b Y)(\lambda\Gamma_b\Gamma_a)^\alpha\lambda^\beta \nonumber\\ 
&=& - (\omega\Gamma^b Y)\Big[ \eta_{ab} \lambda^\alpha \lambda^\beta - 
\frac{1}{2} (\lambda\Gamma_a\Gamma_b)^\alpha\lambda^\beta \Big],
\label{4.48}
\end{eqnarray}
so that also $L^{\alpha\beta\gamma\delta}$  has a residual dependence on 
$Y_\alpha$. This $Y_\alpha$-dependent term in 
$L^{\alpha\beta\gamma\delta}$, which is denoted as 
$L^{\alpha\beta\gamma\delta}|_Y$, reads
\begin{eqnarray}
L^{\alpha\beta\gamma\delta}|_Y &=& \frac{1}{32} \Big[ - (\omega\Gamma_a Y) \lambda^\alpha
\lambda^\beta + \frac{1}{2} (\omega\Gamma^b Y) (\lambda\Gamma_a\Gamma_b)^\alpha\lambda^\beta\Big] 
B^{a\gamma\delta} \nonumber\\ 
&=& \frac{1}{32} \Big[ - (\omega\Gamma_a Y) \lambda^\alpha
\lambda^\beta B^{a\gamma\delta} + {\frac{2}{3}} (\omega\Gamma_b Y) 
(\lambda \Gamma^a\Gamma^b)^\alpha\lambda^\beta  \lambda^\gamma (\Gamma_a\omega)^\delta 
\nonumber\\ 
&+&  \frac{1}{6} (\omega\Gamma_b Y) (\lambda\Gamma^a\Gamma^b)^\alpha\lambda^\beta 
(\lambda\Gamma^a\Gamma^c)^\gamma 
(\Gamma_{c} \omega)^\delta \Big] + \cdots,  
\label{4.49}
\end{eqnarray}
where we have used (\ref{4.46}) with $\omega$ instead of $\tilde\omega$ since 
we have already shown that the dependence on $ Y_\alpha$ in $B_{a}^{\alpha\beta}$ 
drops out.
The first two terms in the r.h.s. of (\ref{4.49}) are $\Gamma_1$-traceless and the third
one vanishes being proportional to the pure spinor constraint by  Fierz
identity. In conclusion the residual $Y_\alpha$-dependence of $L^{\alpha\beta\gamma
\delta}$ does not contribute in $b^{(a)}_4 = L^{\alpha\beta\gamma\delta} 
Z_{\alpha\beta\gamma\delta}$. In fact we have verified explicitly that the 
dangereous terms in $b^{(a)}_4 $, involving $(\tilde\omega\Gamma^{a_1a_2a_3a_4}
\lambda) $ cancel exactly.

Since $L^{\alpha\beta\gamma\delta}$ is defined modulo $ \Gamma_1$-traceless 
terms, one could argue that it should be possible to obtain an expression of
$L$ in terms of  $N^{ab}$ and $J$ only, by adding to (\ref {4.44}) suitable
$\Gamma_1$-traceless terms. We cannot exclude this possibility but all 
attempts we have done thus far were unsuccessful.    

As already remarked, in this section we  worked at the classical level
neglecting quantum corrections from double poles and normal ordering and 
using classical $N^{ab}$, $J$ and $T^{(\omega\lambda)}$ instead of their
exact expressions.
But of course in the final results expressed in terms of  
$N^{ab}$, $J$ and $T^{(\omega\lambda)}$, one should read for these fields
the full covariant and primary fields  (\ref{3.8}),(\ref{3.9}) and
(\ref{3.11}). Moreover these results have to be corrected by adding normal 
order contributions (and perhaps $Y_\alpha$-dependent terms as in 
(\ref{3.8}), (\ref{3.9}) and (\ref{3.11})) in order that the final 
expressions are represented by covariant and primary fields. 
It is important (but not easy) to compute these corrections and verify that 
$b_B$ and $TZ_B$ are primary and covariant operators of ghost number zero 
and that they still obey Eq. (\ref{4.6}) at the quantum level as well.
Without this check, we are not really guaranted that the recipe proposed in 
\cite{Ber6} is completely consistent. 
Unfortunatly, these corrections are not computed in this paper and hopefully will 
be discussed elsewhere.

\section{Conclusion}

In this article, we have constructed a new formalism, which we named 
"Y-formalism", for the pure spinor quantization of superstring theory. 
We have found that although the $\omega$-$\lambda$ OPE is not free owing to 
the presence of the projector $K_\alpha \ ^\beta$, the proper redefinition of 
various operators makes it possible to make a consistent conformal field theory 
with the same OPE's of the Berkovits' covariant pure spinor formalism  
at the quantum level. 

Moreover, using the Y-formalism, we have presented the full expression of the 
picture raised $b_B $ antighost for the first time. Remarkably enough, 
the operator $L^{\alpha\beta\gamma\delta}$  does not seeems to be expressible 
only in terms of the covariant and gauge invariant objects $N^{ab}$ and $J$. 
However, this intrinsic non-covariance of $L^{\alpha\beta\gamma\delta}$ drops out
in the compound object $L^{\alpha\beta\gamma\delta}Z_{\alpha\beta\gamma\delta}$ 
which appears in the $b_B$ antighost owing to the $\Gamma_1$-traceless condition.
Unfortunatly, the expression of $b_B$ is obtained only at the classical level.
It is important to compute the quantum corrections to make sure of the 
consistency of the recipe proposed in \cite{Ber6}. 
In this respect an advantage of the Y-formalism  might appear in the calculations 
relevant to normal ordering and quantum corrections since one has an enough handy 
expression for the  $\omega$-$\lambda$ OPE.
However the discussion of this topic was not done in this paper and is 
postponed to future work.

\begin{flushleft}
{\bf Acknowledgements}
\end{flushleft}

We are grateful to P. Grassi, M. Matone, L.Mazzucato, P. Pasti and D. Sorokin 
for stimulating discussions, and thank N. Berkovits for useful discussions, 
suggestions and comments.
The first author (I.O.) would
like to thank Dipartimento di Fisica, Universita degli Studi di Padova
for its kind hospitality and his work was partially supported by
the Grant-in-Aid for Scientific Research (C) No.14540277 from 
the Japan Ministry of Education, Science and Culture.
 The second author (M.T.) is grateful to the organizers and the participants 
of the nice Workshop on Pure Spinor Formalism at IHES where part of this work
has been presented; his work   was  supported by the European
Community's Human Potential Programme under contract MRTN-CT-2004-005104 
"Constituents, Fundamental Forces and Symmetries of the Universe".

\newpage


\end{document}